\RequirePackage{fix-cm}
\documentclass[twocolumn]{svjour3}          
\smartqed  
\usepackage{graphicx}
\begin{document}

\title{Effect of CuO$_2$ lattice strain on the electronic structure and properties of high-$T_c$ cuprate family}



\author{I.A. Makarov \and
        V.A. Gavrichkov \and
        E.I. Shneyder \and
        I.A. Nekrasov \and
        A.A. Slobodchikov \and
        S.G. Ovchinnikov\and
        A. Bianconi
}


\institute{I.A. Makarov \and V.A. Gavrichkov \and E.I. Shneyder \and S.G. Ovchinnikov \at
              Kirensky Institute of Physics, Federal Research Center KSC SB RAS, 660036 Krasnoyarsk, Russia \\
              \email{maki@iph.krasn.ru}
              \and
I.A. Nekrasov \and A.A. Slobodchikov  \at
              Institute for Electrophysics, Russian Academy of Sciences, Ural Branch, 620016 Ekaterinburg, Russia
              \and
A. Bianconi \at
              Rome International Center for Materials Science Superstripes (RICMASS), 00185 Roma, Italy;CNR-IC, Istituto di Cristallografia, Monterotondo, 00015 Roma, Italy;National Research Nuclear University, MEPhI, 115409 Moscow, Russia
}

\date{Received: date / Accepted: date}

\maketitle

\begin{abstract}
Doping and strain dependences of the electronic structure of the CuO$_6$-octahedra layer within LDA+GTB method in the frameworks of six-band p-d model are calculated. Band structure and Fermi surface of the quasiparticle excitations in the effective Hubbard model are characterized by inhomogeneous distribution of the ${\bf{k}}$-dependent spectral weight. Doping results in reconstruction of the band structure, redistribution of the spectral weight over dispersion surface and reconstruction of Fermi surface from small hole pockets in the underdoped region to large hole contour in the overdoped region through two quantum phase transitions (QPT). Strain increasing leads to displacement of the valence and conductivity bands, bandwidths decreasing and shift of the concentrations corresponding to QPTs. Strain dependences of the antiferromagnetic exchange coupling and DOS at the Fermi level determining superconducting temperature ${T_c}$ are obtained. Effective exchange coupling in the equation for ${T_c}$ demonstrates monotonic strain dependence.

\keywords{Cuprate superconductors \and CuO$_2$ lattice strain \and Strong electronic correlations \and Hubbard model \and Electronic structure \and Lifshitz transitions}
\end{abstract}

\section{Introduction}
\label{sec_introduction}
It is wide-spread believe that all hole-doped high-temperature superconducting cuprates have common generic phase diagram in the plane (concentration of doped holes $x$ per Cu site, temperature $T$). In fact, the phase diagram of cuprates is determined not only by doping degree $x$ but also by the strain of CuO$_2$ lattice. Many evidences of the influence of the strain of CuO$_2$ layer on superconducting and competing phases were obtained in the undoped and doped systems \\(La$_{1-y}$Ln$_y$)$_{2-x}$M$_x$CuO$_{4+\delta}$ (Ln$^{3+}$ is lanthanides and M$^{2+}$ is elements Ca, Sr, Ba) in the experiments using epitaxial strain and cation substitutions in the rocksalt spacers with the same valence but different radii. It was shown that in La$_2$CuO$_4$ film compressive strain induced by the lattice mismatch with the substrate perovskite film results in significant $T_c$ increasing~\cite{Arrouy1996,Sato1997,Lockuet1998,Sato2000}. Lattice misfit strain between the CuO$_2$ planes and the intercalated rocksalt layers determines variation $T_c$ within family of (Ln$_{1-y}$M$_y$)$_2$CuO$_4$ superconductors~\cite{Attfield1998} and different maxima of the critical temperature in the different cuprate families~\cite{AgrBB2003}. Two critical strains in the oxygen doped La$_2$CuO$_{4+\delta}$ and Nd doped La214 system were discovered with EXAFS spectra~\cite{BiancSaiAgr2000} and x-ray diffuse scattering~\cite{Kusmartsev2000,BiancCastBianc2000}: first critical strain determines onset of local lattice deformations and short-range striped domains formation, long-range charge ordering (insulating crystal of polaronic strings~\cite{Kusmartsev2000}) at doping $\delta=1/8$ appears at strain above the second critical value. Critical value of buckling of CuO$_2$ layer (Cu-O-Cu angle) determines boundaries of insulating antiferromagnetic and superconducting phases inside low-temperature tetragonal striped phase in the undoped parent compound (La$_{1-y}$Nd$_y$)$_{2-x}$Sr$_x$CuO$_4$~\cite{Buchner1994}. 

One can imagine that all material aspects may be solved explicitly by the density functional theory (DFT) as it happens now in many condensed materials. Nevertheless cuprates and many other so called strongly correlated materials cannot be treated successfully by the DFT. That is why more simple model approaches are involved to understand the material dependent properties of cuprates. The importance of strong electron correlations in the CuO$_2$ planes was pointed out soon after the discovery of superconductivity in cuprates~\cite{Anderson1987}. Following the discovery that the electronic holes in the CuO$_2$ induced by chemical doping are injected in the O(2p) orbitals~\cite{Bianconi1987} the microscopic analysis of the electronic orbitals of copper and oxygen with account for strong correlations results in the three-band Hubbard model, or the p-d model for CuO$_2$ square lattice ~\cite{Emery1987}. The single-band Hubbard model for CuO$_2$ planes is the most known, it may be deduced from the general multielectron approach as the effective low energy model for undoped and weakly doped cuprates~\cite{KorGavOvch2004,KorGavOvch2005}.

In this paper we will discuss material dependent electronic properties within the effective Hubbard model with parameters obtained from the DFT-LDA calculations. Additional material parameter will be the isotropic lattice strain of the CuO$_6$-octahedron. We start with the plane of CuO$_6$-octahedra in undoped La$_2$CuO$_4$ and calculate its lattice parameter ${a_0}$. For other cuprates we will considered also the CuO$_2$ planes with material specific lattice parameter $a$, and the lattice strain will be determined by ${{\delta a} \mathord{\left/
 {\vphantom {{\delta a} {{a_0} = {{\left( {a - {a_0}} \right)} \mathord{\left/
 {\vphantom {{\left( {a - {a_0}} \right)} {{a_0}}}} \right.
 \kern-\nulldelimiterspace} {{a_0}}}}}} \right.
 \kern-\nulldelimiterspace} {{a_0} = {{\left( {a - {a_0}} \right)} \mathord{\left/
 {\vphantom {{\left( {a - {a_0}} \right)} {{a_0}}}} \right.
 \kern-\nulldelimiterspace} {{a_0}}}}}$, and may be positive as well as negative. Similar model has been used for the analysis of the experimental data for different cuprates and construct the phase diagram in the plane (doping, strain). Here we will calculate the doping and strain dependent electronic structure in the normal state, and discuss the material dependence of the Lifshitz transitions with the change the Fermi surface topology under doping, the density of states at the Fermi level $N\left( \mu  \right)$, and the interatomic exchange coupling parameter $J$. We will study later the effect of in-plane anisotropic misfit strain on the electronic structure of the CuO$_2$ plane giving an anisotropic distortion of the CuO$_6$ octahedra, with both rotation of the CuO$_4$ square planes and anisotropic compression of the Cu-O bond lengths observed at optimum doping in cuprates~\cite{Bianconi1996}.

The paper is organized as follows: in part 2 we will discuss the ab initio calculation of strain dependent model parameters, in part 3 the doping dependent electronic structure and Fermi surface without strain are discussed, in part 4 we analyze the strain dependence of electronic structure. In part 5 the conclusion is given.

\section{Strain dependence of the six-band p-d model parameters within the LDA-GTB approach}
\label{LDAGTB}

\begin{figure}
\center
\includegraphics[width=0.6\linewidth]{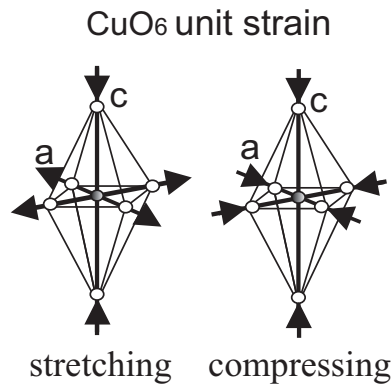}
\caption{\label{fig:CuO} Microscopic simulation for the CuO$_6$-octahedron strain with $\delta a/a_0\neq0$ induced by in-plane stretching (left side) and compressing (right side) CuO$_6$-octahedron. An elongation of c-parameter was derived from an empirical condition of the constant unit cell volume $V(\delta a/a_0)=V_0$}
\end{figure}

\begin{table*} [htb]
\center
\begin{tabular}{c@{\hspace{5mm}}c@{\hspace{8mm}}*{12}{c@{\hspace{8mm}}}}
$\delta a/a_0$ & $-1 \%$ & $-0.5 \%$ & $0 \%$ & $0.5 \%$ & $1.5 \%$ & $2.5 \%$ & $3.5 \%$ & $4.15 \%$ \\
\hline
  ${E_{{x^2}}}$ & -1.79 &	-1.824 &	-1.861 &	-1.9 &	-1.657 &	-2.047 &	-2.124 &	-2.18 \\
  ${E_{{z^2}}}$ & -2.056	& -2.075	& -2.097	& -2.119	& -1.821	& -2.186	& -2.227	& -2.260 \\
  ${E_{{p_x}}}$ & -2.724	& -2.775	& -2.825	& -2.863	& -2.541	& -2.980	& -3.026	& -3.053 \\
  ${E_{{p_y}}}$ & -2.724	& -2.775	& -2.825	& -2.863	& -2.541	& -2.980	& -3.026	& -3.056 \\
  ${E_{{p_z}}}$ & -1.741	& -1.727	& -1.721	& -1.720	& -1.541	& -1.690	& -1.713	& -1.729 \\
  $t\left( {{d_x},{p_x}\left( {{p_y}} \right)} \right)$ & 1.450	& 1.427	& 1.403	& 1.379	& 1.313	& 1.280	& 1.232	& 1.201 \\
  $t\left( {{d_z},{p_x}\left( {{p_y}} \right)} \right)$ & 0.5170	& 0.5200	& 0.5230	& 0.5260	& 0.5340	& 0.5490	& 0.5620	& 0.5700 \\
  $t\left( {{p_x},{p_y}} \right)$ & 0.8930	& 0.8760	& 0.8590	& 0.8420	& 0.8090	& 0.7680	& 0.7320	& 0.7100 \\
  $t\left( {{d_z},{p_z}} \right)$ & 0.7790	& 0.7980	& 0.8200	& 0.8380	& 0.8600	& 0.8800	& 0.9050	& 0.9180\\
  $t\left( {{p_x}\left( {{p_y}} \right),{p_z}} \right)$ & 0.3790	& 0.3910	& 0.4030	& 0.4150	& 0.4280	& 0.4430	& 0.4560	& 0.4620 \\
\end{tabular}
\caption{Values of on-site energies and hopping integrals (in eV) for the tetragonal La$_2$CuO$_4$ obtained during projection on the Wannier function in the frameworks of the six-band model. Here, ${x^2}$, ${z^2}$, ${p_x}$, ${p_y}$, ${p_z}$ denote Cu-${d_{{x^2} - {y^2}}}$, Cu-${d_{3{z^2} - {r^2}}}$, O-${p_x}$, O-${p_y}$, O-${p_z}$ electronic orbitals, respectively.}
\label{Param}
\end{table*}

\begin{figure*}
\center
\includegraphics[width=0.49\linewidth]{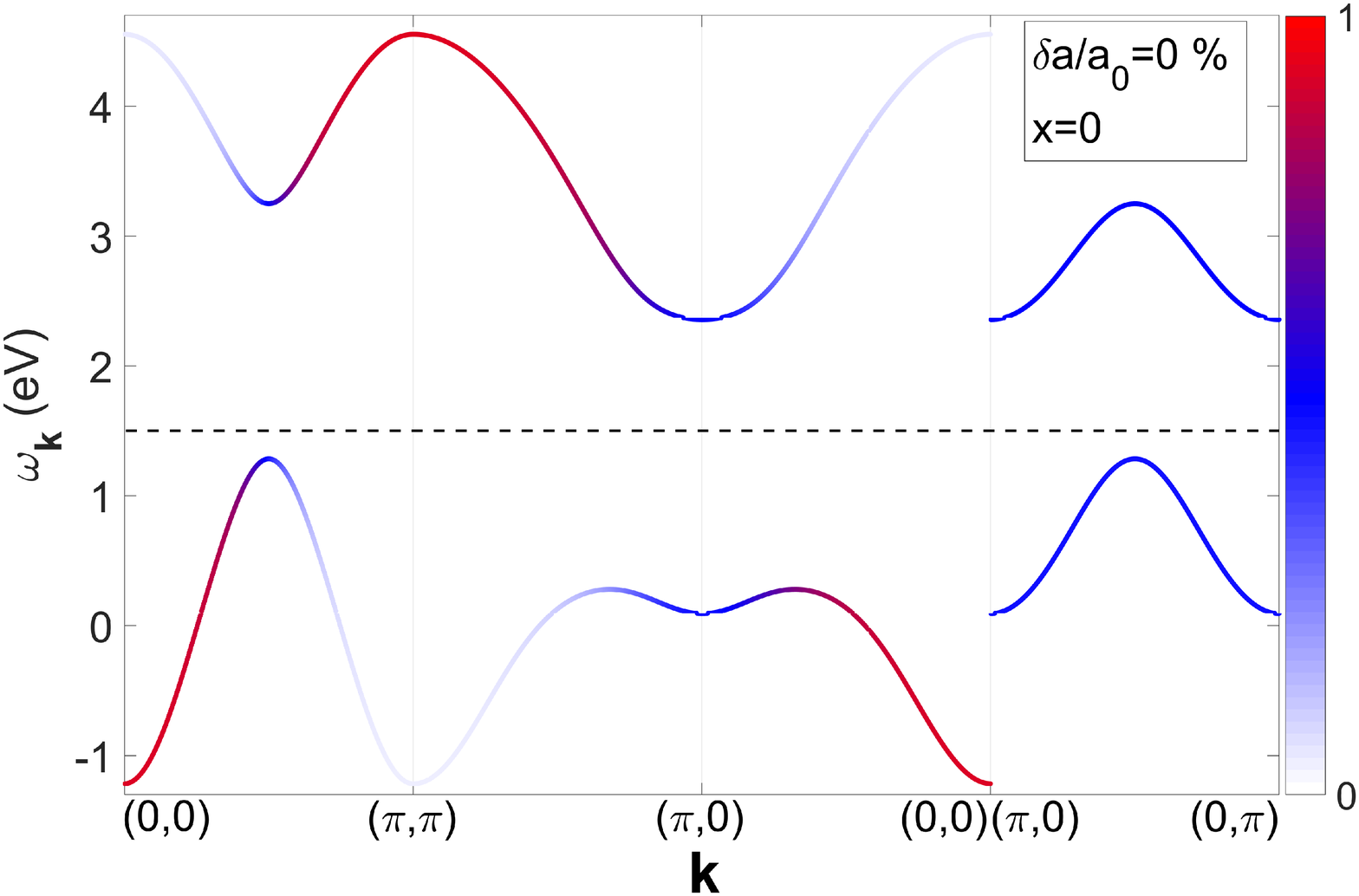}
\includegraphics[width=0.49\linewidth]{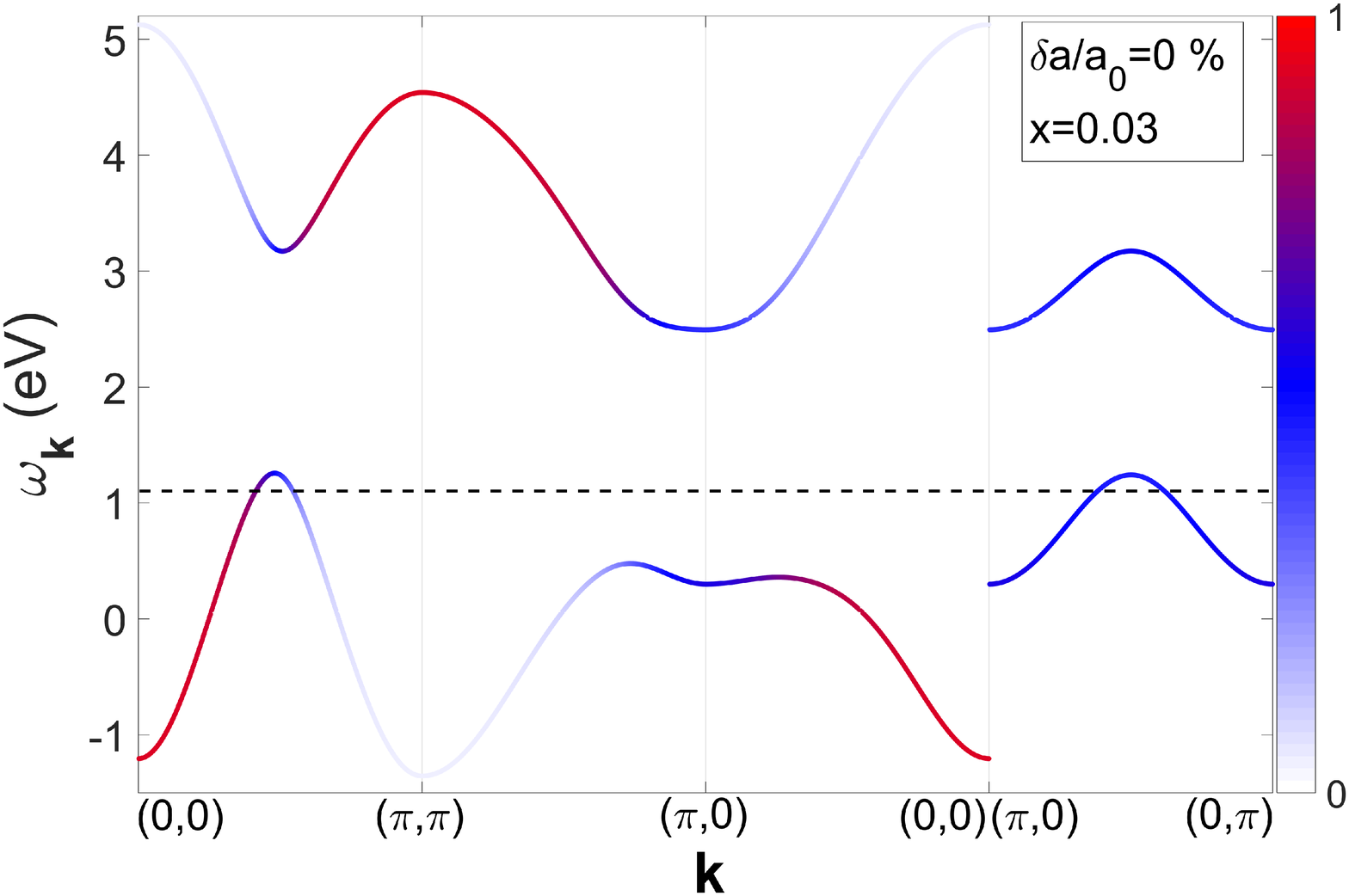}
\includegraphics[width=0.49\linewidth]{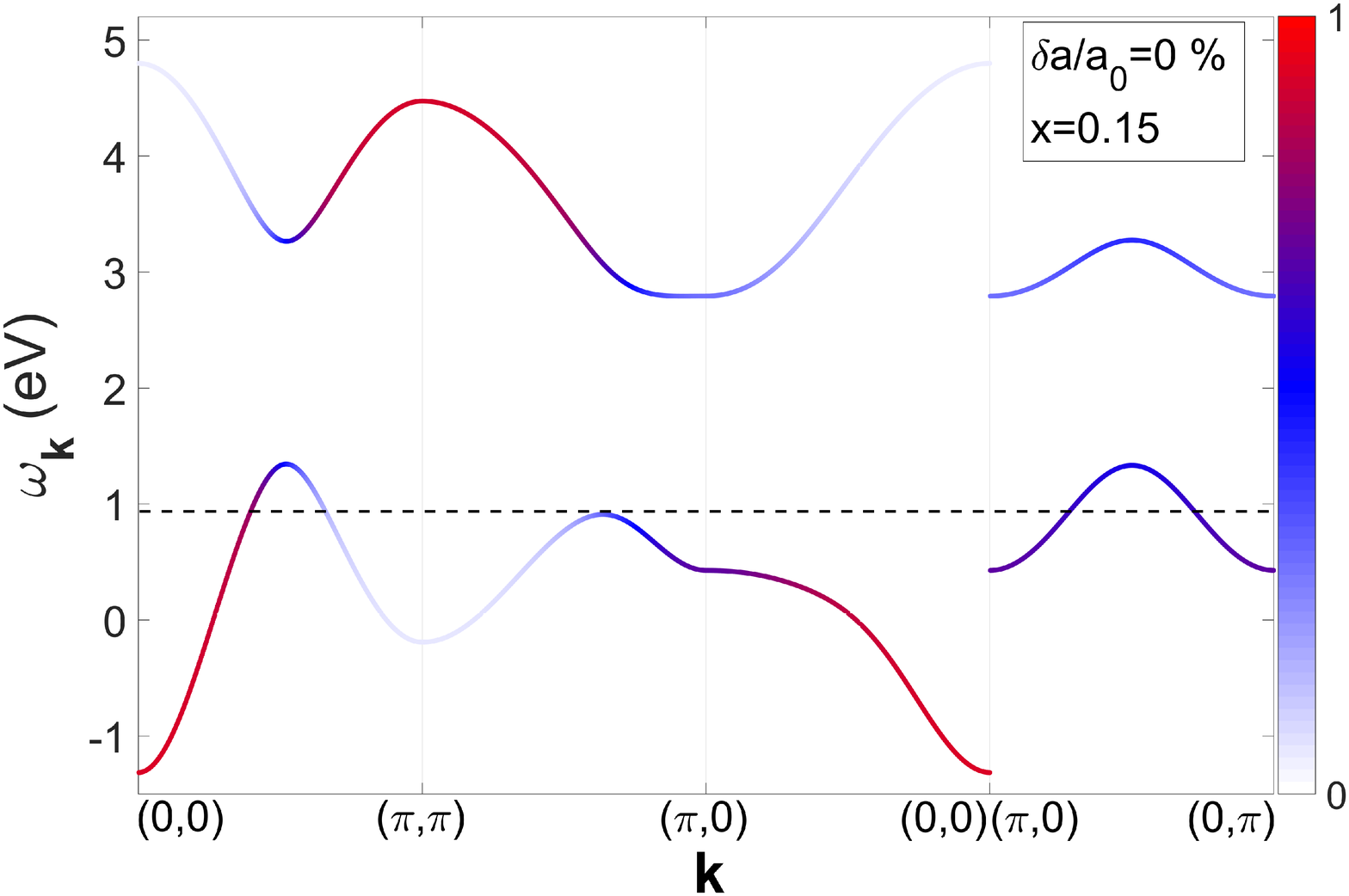}
\includegraphics[width=0.49\linewidth]{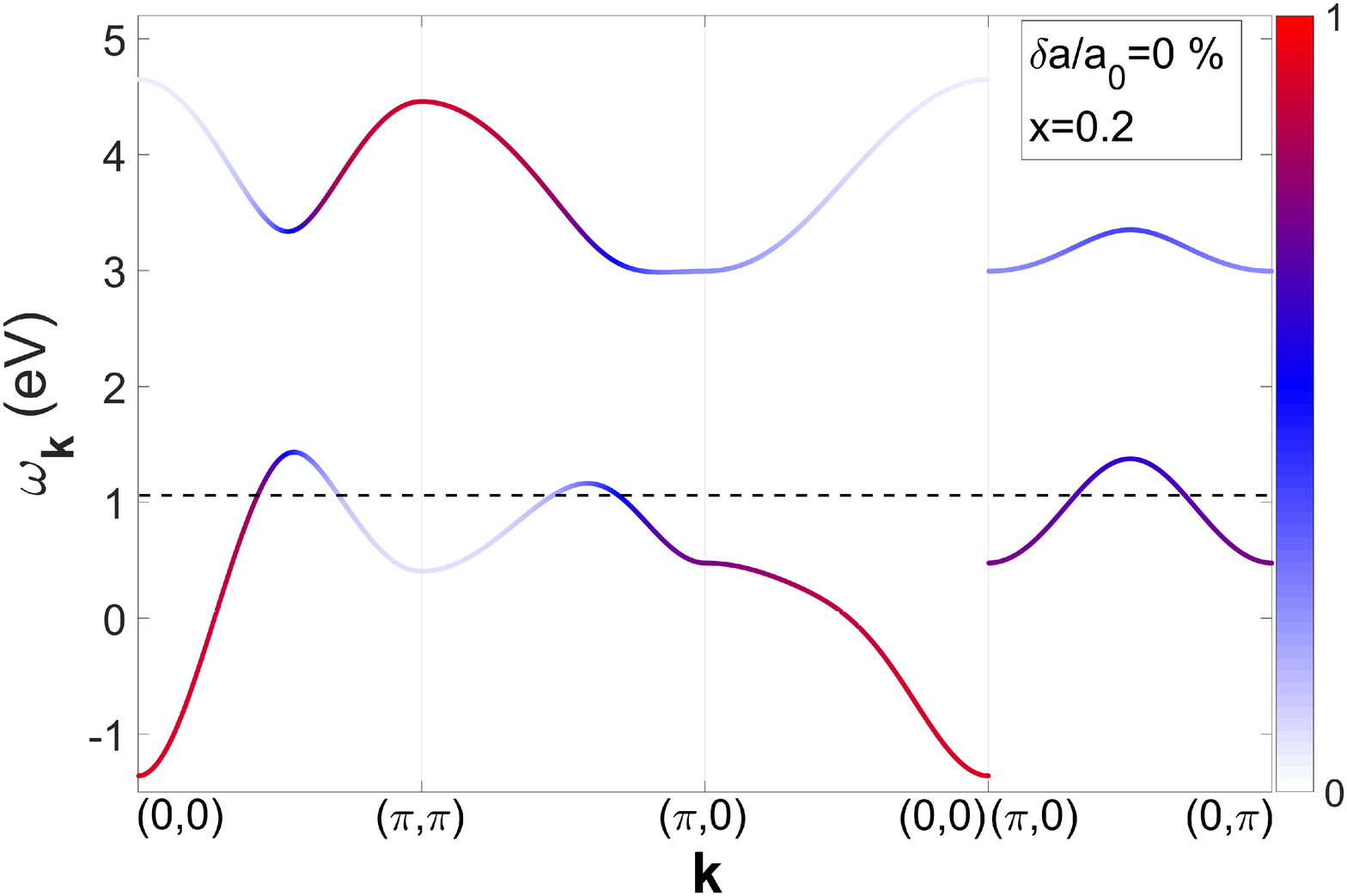}
\includegraphics[width=0.49\linewidth]{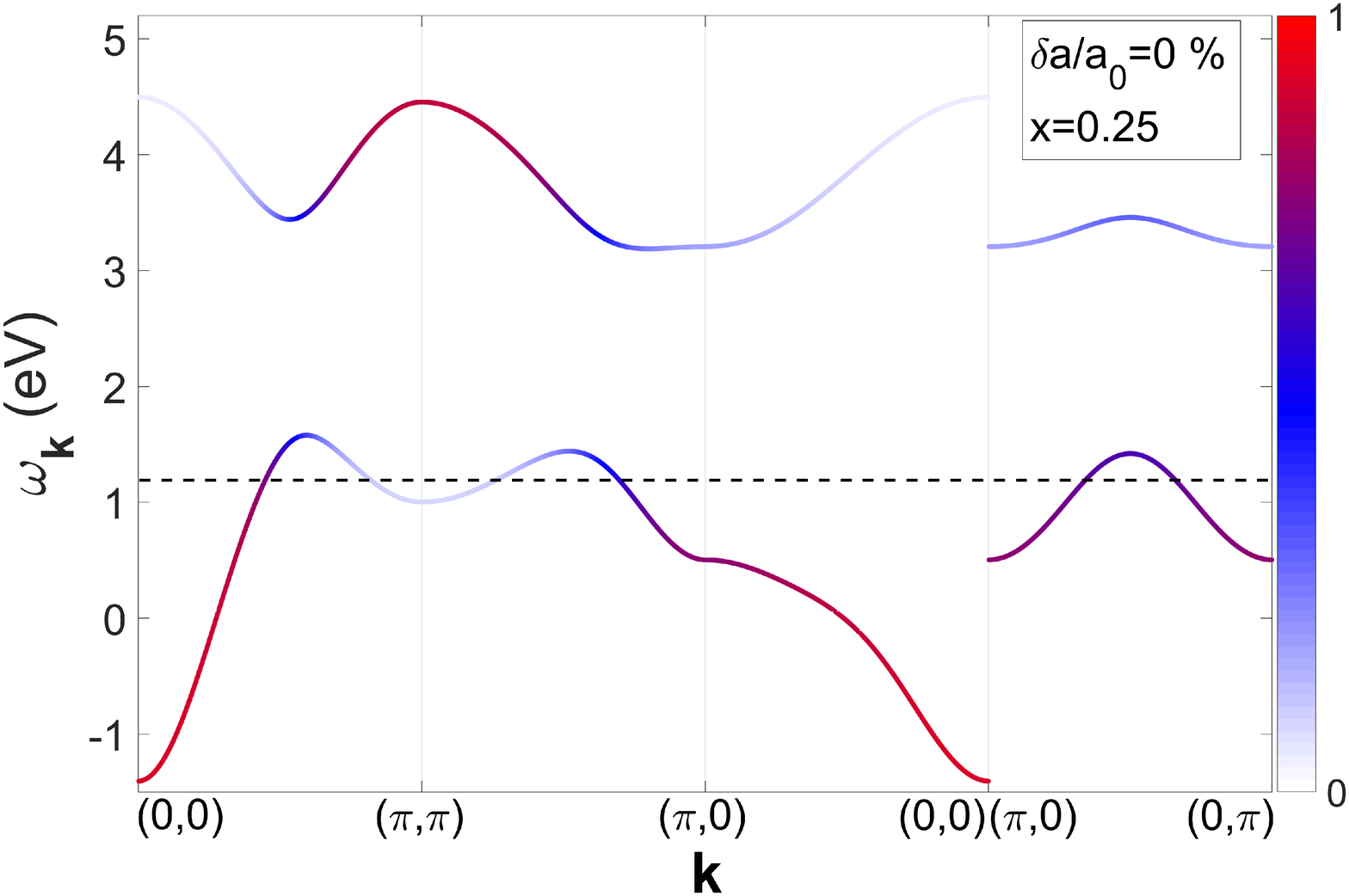}
\includegraphics[width=0.49\linewidth]{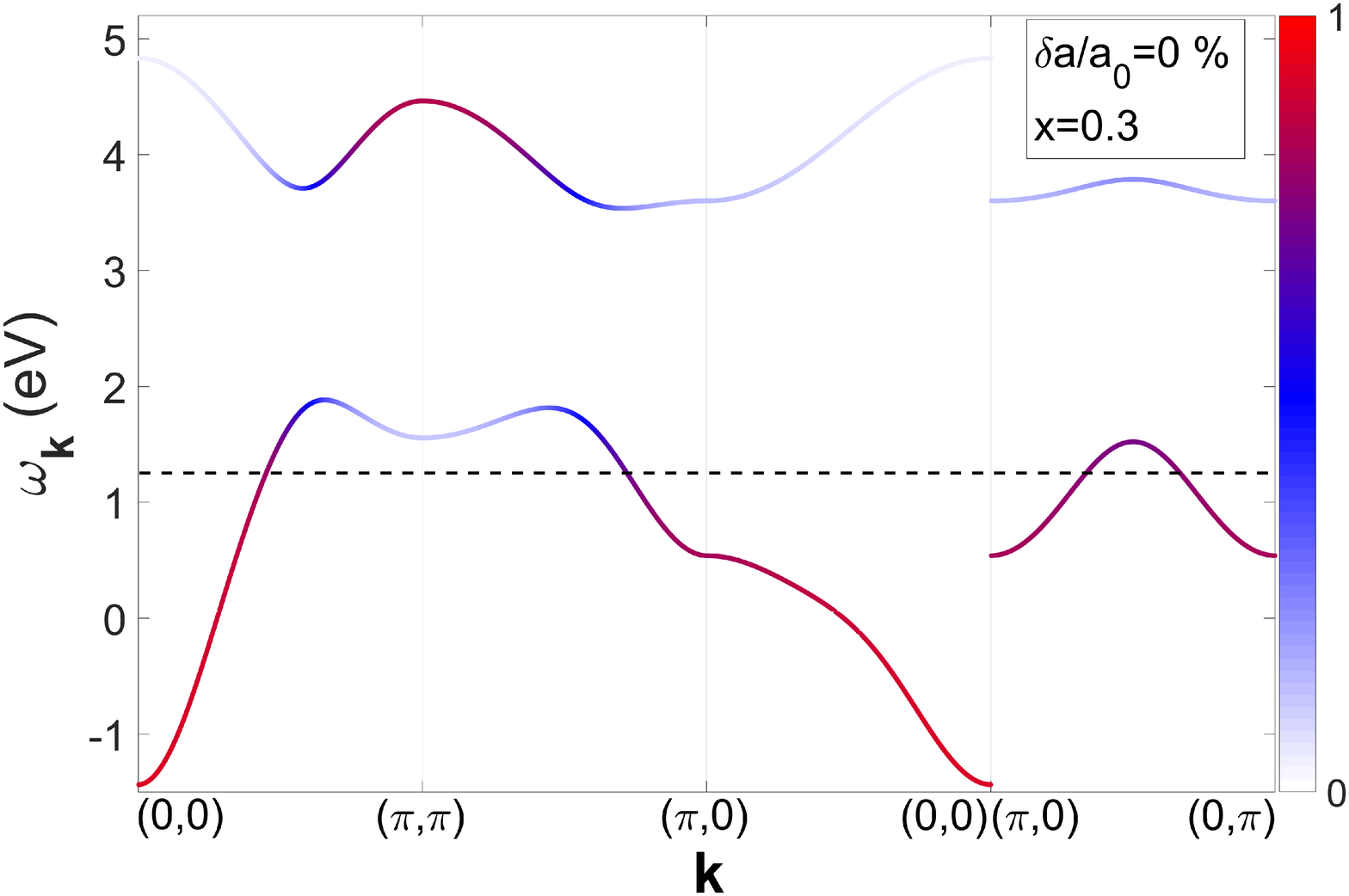}
\caption{\label{fig:BandStr} (Color online) (a)-(f) Reconstruction of the Hubbard fermions bands with hole doping at zero strain $\delta a/a_0=0 \%$. Color of the each point of the dispersion curve indicates quasiparticle spectral weight at given $k$-point. Dashed line indicates the Fermi level}
\end{figure*}

We are interested in the study of strain effect on characteristics of electronic structure related to superconducting phase therefore it is important to describe properly states with energies near chemical potential. Chemical potential is lowered approximately by $1$ eV into the LHB with doping up to overdoped region. Electronic structure of La$_{2-x}$Sr$_x$CuO$_4$ compound in the interesting energy interval of the LHB will be described by six-band p-d model for one layer of CuO$_6$-octahedra. This model includes ${d_{{x^2} - {y^2}}}$-, ${d_{3{z^2} - {r^2}}}$-orbitals of planar Cu atoms, ${p_{x,y}}$-orbitals of planar O atoms and ${p_z}$-orbitals of two apical O atoms in each CuO$_6$-octahedron. Six-band p-d model takes into account on-site energies of all considered orbitals, hopping between them, intra- and interatomic Coulomb interaction and exchange interaction. Calculation of electronic structure will be realized within LDA+GTB method~\cite{KorGavOvch2005}. LDA method provides parameters of the model Hamiltonian. Generalized tight-binding (GTB) method allows to construct the effective Hamiltonian for quasiparticle excitations in the terms of the Hubbard operators, this method is actual for description of systems with strong electronic correlations.

On-site energies and hopping integrals can be calculated using the procedure of LDA eigenfunctions projection onto Wannier functions basis of the chosen model. The calculation was carried out using the Linearized Muffin-Tin Orbitals method (LMTO)~\cite{Andersen1975,GunnJepAnd1983,AndJep1984} with some modifications~\cite{Amulet}. The $k$-points grid is 12x12x12. The lattice parameters for undistorted La$_2$CuO$_4$ in the tetragonal phase are ${a_0} = {b_0} = 3.783{\AA}$, ${c_0} = 13.2883{\AA}$. The strain effect of CuO$_6$-octahedron is simulated by the set of deformations ${{\delta a} \mathord{\left/
 {\vphantom {{\delta a} {{a_0}}}} \right.
 \kern-\nulldelimiterspace} {{a_0}}}$. The lattice parameter $a$ was changed by $-1\%$, $-0.5\%$, $+0.5\%$, $+1.5\%$, $+2.5\%$, $+3.5\%$, $+4.15\%$ in-plane compressing and stretching respectively. Accordingly parameter $c$ was calculated under the condition of constant CuO$_6$-octahedron volume. The volume of the whole unit cell is also preserved. Table~\ref{Param} below shows values of the on-site energies and hopping parameters at different strains. Since the relation between the in-plane compressing (stretching) and microscopic picture is empirical, we simulate the strain effect at ${{\delta a} \mathord{\left/
 {\vphantom {{\delta a} {{a_0} \ne 0}}} \right.
 \kern-\nulldelimiterspace} {{a_0} \ne 0}}$ within the approach (see Fig.~\ref{fig:CuO}). Coulomb and exchange parameters were obtained in constrained LDA supercell calculations~\cite{GunnAndJepZaa1989,AnGunn1991}.

Coulomb repulsion on one site results in the dependence of electron (or hole) energy on occupation of the site on which this electron is located. In this case there are two types of particles on each site: (i) electron added to the empty site and (ii) electron added to the site that is occupied by electron with the opposite spin projection. Description of such system can be performed using representation of electron as superposition of Fermi-type excitations between different multielectron initial and final states. Generalized tight-binding (GTB) method~\cite{OvchSan1989,GavOvchBorGor2000} is realization of this representation of quasiparticle excitations. Each quasiparticle excitation is defined as transition between multiparticle initial and final states of the single cluster, it acquires dispersion due to intercell hopping. Electronic structure in the crystal lattice is formed by bands of quasiparticle excitations. Thus GTB method is the cluster form of perturbation theory in the terms of quasiparticle excitations. In this work cluster is chosen as single CuO$_6$-octahedron. First step of GTB method is representation of full Hamiltonian in the form of sum of Hamiltonian of intracluster interactions and Hamiltonian of intercluster interactions $H = \sum\limits_f {H_f^c}  + \sum\limits_{fg} {H_{fg}^{cc}}$. Second step is exact diagonalization of Hamiltonian $H_f^c$ of the cluster $f$ with different number of carriers (in cuprates of p-type carriers are holes). The result of exact diagonalization is a set of multihole local cluster eigenstates. To reproduce electronic structure in the upper part of the valence band and the lower part of the conductivity band it is sufficient to consider cluster with number of holes ${n_h} = 0,1,2$ and to take into account only ground cluster eigenstates. Eventually we come to basis of the effective Hubbard model: zero-hole state $|0\rangle $ (hole vacuum, electronic configuration ${d^{10}}{p^6}$), single-hole states $\left| \sigma  \right\rangle $ and $\left| {\bar \sigma } \right\rangle $ (superpositions of the configurations ${d^9}{p^6}$ and ${d^{10}}{p^5}$) degenerate in the paramagnetic phase, and two-hole singlet state $\left| S \right\rangle $ (mix of the configurations ${d^8}{p^6}$, ${d^9}{p^5}$ and ${d^{10}}{p^4}$). Doping by holes increases occupation of two-hole cluster eigenstate $\left| S \right\rangle $ and decreases occupation of single-hole cluster eigenstates $\left| \sigma  \right\rangle $ and $\left| {\bar \sigma } \right\rangle $.

Third step is the construction of the quasiparticle excitations which are transitions between multihole cluster eigenstates with the number of holes differs by one. These Fermi-type quasiparticle excitations are called Hubbard fermions. There are four quasiparticle excitations in the Hubbard model: excitations between zero-hole state $\left| 0 \right\rangle $ and single-hole states ($\left| \sigma  \right\rangle $, $\left| {\bar \sigma } \right\rangle $) form upper Hubbard band (UHB) of electrons or the conductivity band (CB). Excitations between states $\left| \sigma  \right\rangle $, $\left| {\bar \sigma } \right\rangle $ and two-hole states ($\left| S \right\rangle $) form lower Hubbard band (LHB) of  electrons or the valence band (VB). Each excitation is described by the Hubbard operator ${X^{pq}} = \left| p \right\rangle \left\langle q \right|$. At the last step the total Hamiltonian is rewritten in the terms of Hubbard operators.
To obtain dispersion of Hubbard fermions we use equation of motion for the matrix Green function $\hat D\left( {f,g;t,t'} \right)$  with elements ${D_{mn}}\left( {f,g;t,t'} \right) = \left\langle {\left\langle {{X_f^m\left( t \right)}}
 \mathrel{\left | {\vphantom {{X_f^m\left( t \right)} {\mathop {X_g^n\left( {t'} \right)}\limits^\dag  }}}
 \right. \kern-\nulldelimiterspace}
 {{\mathop {X_g^n\left( {t'} \right)}\limits^\dag  }} \right\rangle } \right\rangle $, where $m,n$ are indexes of the quasiparticle excitations, each index is uniquely defined by initial and final states of excitation $m \equiv \left( {p,q} \right)$. Equation of motion is decoupled by applying the Mori-type projection technique to the matrix Green function $\hat D\left( {f,g;t,t'} \right)$~\cite{KuzOvch1977}. The Dyson equation for matrix Green function $\hat D\left( {{\bf{k}};\omega } \right)$ in the momentum space looks like~\cite{OvchVal2004}:

\begin{equation}
\label{Dysoneq}
\hat D\left( {{\bf{k}};\omega } \right) = {\left[ {\omega \hat E - \hat \Omega  - \hat F\hat t\left( {\bf{k}} \right) - \hat \Sigma \left( {{\bf{k}};\omega } \right)} \right]^{ - 1}}\hat F
\end{equation}
Here $\hat E$ is the unit matrix, $\hat \Omega $ is diagonal matrix of local quasiparticle energies with matrix elements $\Omega \left( m \right) = \Omega \left( {pq} \right) = {\varepsilon _p} - {\varepsilon _q}$, where ${\varepsilon _p}$ is the energy of the cluster eigenstate $p$. $\hat F$ is diagonal matrix of the filling factors of the quasiparticles, its diagonal matrix elements are $F\left( m \right) = F\left( {pq} \right) = \left\langle {{X^{pp}}} \right\rangle  + \left\langle {{X^{qq}}} \right\rangle $. Filling numbers of cluster eigenstates $\left\langle {{X^{pp}}} \right\rangle $ is determined self-consistently from the condition of local basis completeness $\sum\limits_{{n_h} = 0}^2 {\sum\limits_p {X_f^{pp}} }  = 1$ and the chemical potential equation $n = 1 + x = \sum\limits_{{n_h} = 0}^2 {\sum\limits_p {{n_h}\left\langle {X_f^{pp}} \right\rangle } }$ (here $n = 1 + x$ is the hole concentration for La$_{2-x}$Sr$_x$CuO$_4$). $\hat t\left( {\bf{k}} \right)$ is the matrix of intercluster hoppings, its matrix elements are defined in work~\cite{KorGavOvch2005}. $\hat \Sigma \left( {{\bf{k}};\omega } \right)$ is the self-energy matrix which contains spin-spin and kinematic correlation functions. In a static limit of the self-energy we obtain the generalized mean field approximation as has been described in~\cite{PlAnAdAd2003}. Spin-spin correlation functions were taken from work~\cite{KorOvch2007}. Kinematic correlation functions are calculated self-consistently with filling numbers and chemical potential.

\section{Doping dependence of the electronic structure in the system without strain \label{Dopdep_without_strain}}

\begin{figure}
\center
\includegraphics[width=0.98\linewidth]{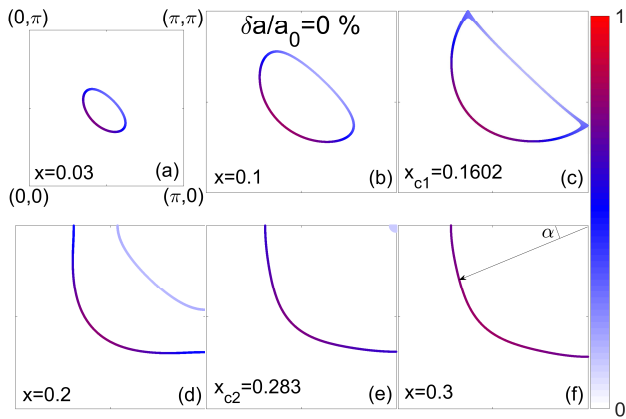}
\caption{\label{fig:FS} (Color online) (a)-(f) Evolution of Fermi surface with doping at zero strain $\delta a/a_0=0 \%$ from small hole pocket in the underdoped compound to large hole Fermi contour in overdoped compound through two quantum phase transitions at ${x_{c1}}$ and ${x_{c2}}$. Subfigure (f) contains definition of Fermi angle $\alpha $}
\end{figure}

Wide valence band with homogeneous distribution of the spectral weight over all $k$-points with minimum at ${\bf{k}} = \left( {0,0} \right)$ and maximum at ${\bf{k}} = \left( {\pi ,\pi } \right)$ obtained in LDA calculations is described by dispersion of single electron within tight-binding (TB) model. In the system with strong Coulomb correlations single electron band is splitted into two Hubbard subbands, width of each subband is smaller than of the free electron band in TB model. Without interband hopping the spectral weight of electrons in these subbands is homogeneous at all k-points and twice smaller than of electron band in TB model.  Interband hoppings mix two types of electrons (i) and (ii), VB (and CB) is formed by superposition of LHB and UHB excitations. Band dispersion is renormalized and distribution of spectral weight is inhomogeneous in the VB and CB now.  Spin-spin and kinematic correlations result in additional reconstruction of subbands.

Undoped HTSC cuprates are antiferromagnetic insulators. Band structure at $x = 0$ is obtained with separate calculation (Fig.~\ref{fig:BandStr}a). To describe long range antiferromagnetic state at finite temperatures we consider two sublattices~\cite{MakOvch2015}. For doped system we consider paramagnetic spin liquid state with spin-spin correlation functions which have short-range antiferromagnetic type. Fig.~\ref{fig:BandStr}(b)-(f) shows band structure of Hubbard fermions in the hole doped systems. Spin correlations suppresses energy of quasiparticles with ${\bf{k}} = \left( {\pi ,\pi } \right)$ in the VB and enhances energy of quasiparticles with ${\bf{k}} = \left( {0,0} \right)$ in the CB. Spectral weight is inhomogeneously distributed over dispersion surface throughout the whole Brillouin zone. Spectral weight in the VB is maximal at point ${\bf{k}} = \left( {0,0} \right)$, it decreases with increasing ${\bf{k}}$ up to the boundaries of Brillouin zone. Inverse tendency takes place for the CB, maximal spectral weight is at point ${\bf{k}} = \left( {\pi ,\pi } \right)$ and minimal at ${\bf{k}} = \left( {0,0} \right)$. Both VB and CB are formed by high-intensity and low-intensity parts which are connected at points ${\bf{k}} = \left( {{\pi  \mathord{\left/
 {\vphantom {\pi  2}} \right.
 \kern-\nulldelimiterspace} 2},{\pi  \mathord{\left/
 {\vphantom {\pi  2}} \right.
 \kern-\nulldelimiterspace} 2}} \right)$ and at $k$-point in the direction ${\bf{k}} = \left( {\pi ,0} \right)$ - ${\bf{k}} = \left( {\pi ,\pi } \right)$. The general structure of bands can be presented as result of hybridization of high-intensity single band of free electrons in antiferromagnetic lattice and shadow low-intensity band~\cite{NekKuchPchSad2007,Nekrasov2009}. It is seen that band structures at $x = 0$ and at $x = 0.03$ are identical except for slight difference in quasiparticle energy at the points ${\bf{k}} = \left( {0,0} \right)$ and ${\bf{k}} = \left( {\pi ,\pi } \right)$. Note band structure of quasiparticle excitations with spin projection "up" (+1/2) and spin projection "down" (-1/2) coincides.

\begin{figure}
\center
\includegraphics[width=0.98\linewidth]{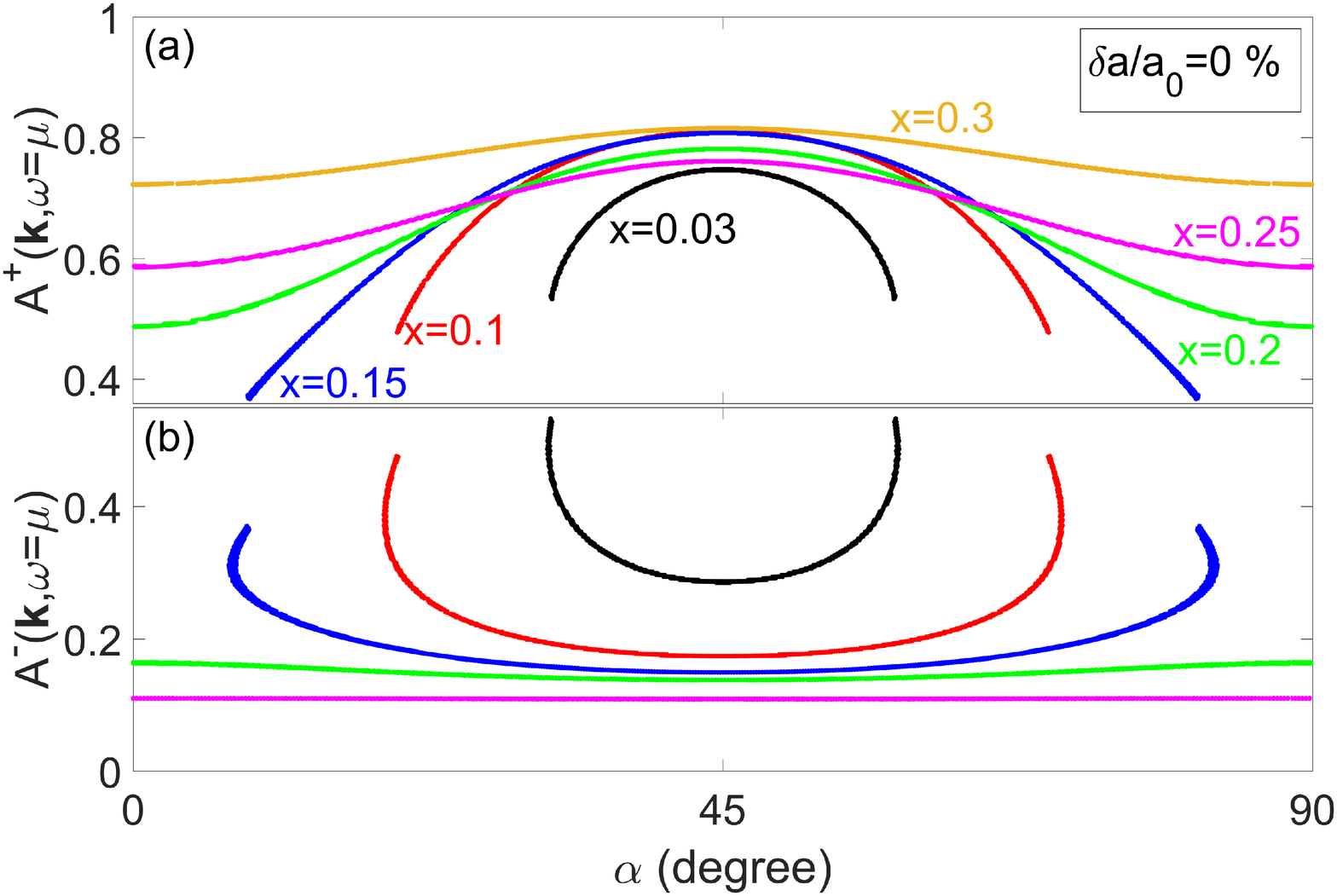}
\includegraphics[width=0.8\linewidth]{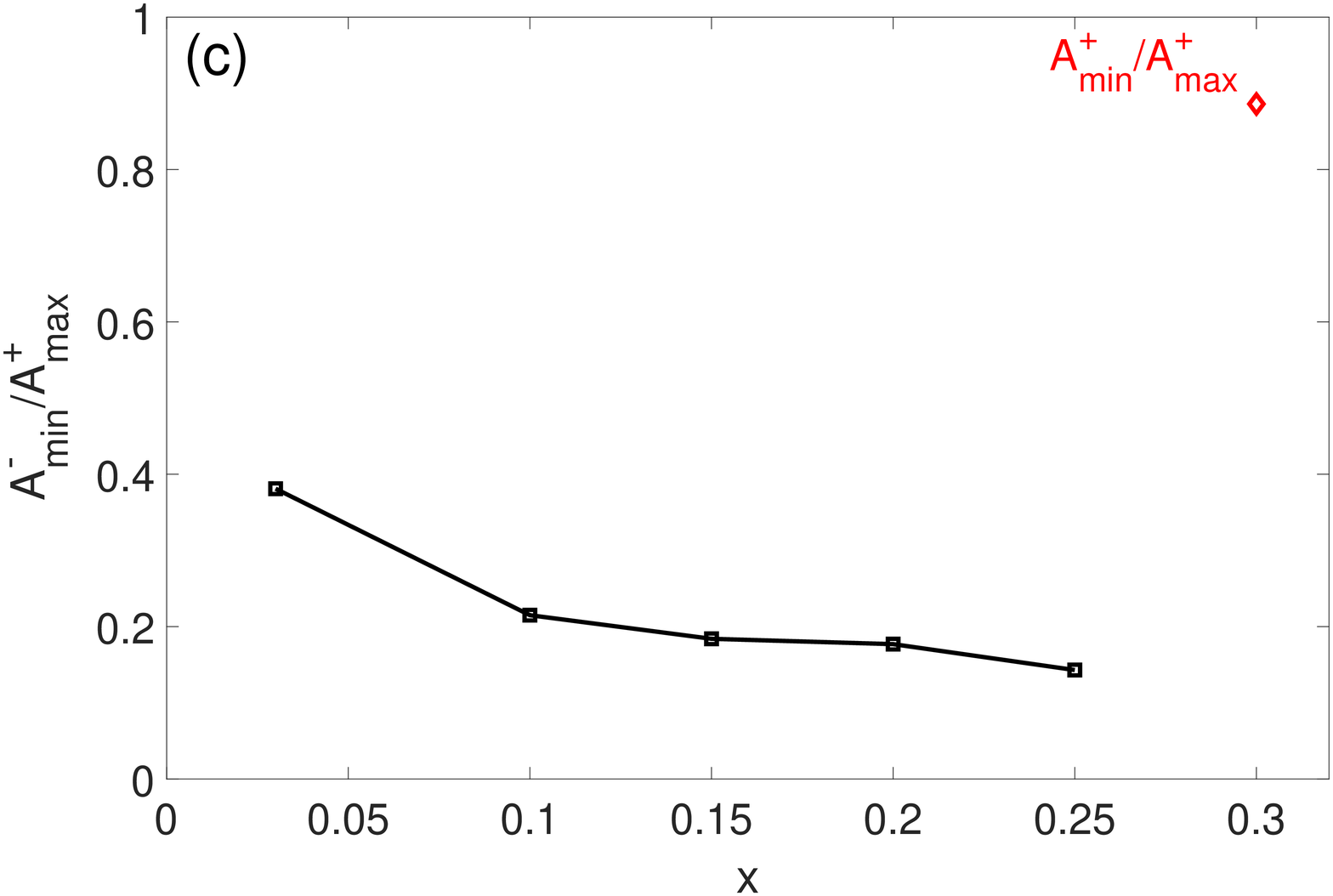}
\caption{\label{fig:Inhom} (Color online) (a) Intensity profile along the high-intensity and low-intensity sides of Fermi contour as a function of Fermi angle $\alpha $ (defined in Fig.~\ref{fig:FS}f) for various hole concentrations. (b) Concentration dependence of the parameter of the spectral weight inhomogeneity $\Delta A=A_{min}^{-}/A_{max}^{+}$. There is only one Fermi contour at hole concentration $x = 0.3$ so parameter of the spectral weight inhomogeneity characterizes ratio between maximal and minimal spectral weight on this hole contour}
\end{figure}

At small concentration of doped holes ($x = 0.03$) the Fermi contour (FC) is a small hole pocket centered around ${\bf{k}} = \left( {{\pi  \mathord{\left/
 {\vphantom {\pi  2}} \right.
 \kern-\nulldelimiterspace} 2},{\pi  \mathord{\left/
 {\vphantom {\pi  2}} \right.
 \kern-\nulldelimiterspace} 2}} \right)$ (Fig.~\ref{fig:FS}a). Spectral weight is inhomogeneously distributed over FC for all doping levels (Fig.~\ref{fig:FS}(a)-(f)). Maximal spectral weight is in the nodal direction on the one side of hole pocket close to ${\bf{k}} = \left( {{\pi  \mathord{\left/
 {\vphantom {\pi  2}} \right.
 \kern-\nulldelimiterspace} 2},{\pi  \mathord{\left/
 {\vphantom {\pi  2}} \right.
 \kern-\nulldelimiterspace} 2}} \right)$), on the opposite side of pocket spectral weight is suppressed. This is well seen on the angular profile of spectral weight distribution over high-intensity ${A^ + }$ (Fig.~\ref{fig:Inhom}a, black line) and low-intensity ${A^ - }$ (Fig.~\ref{fig:Inhom}b, black line) sides of the hole pocket. The intensity of the quasiparticle excitations monotonically decreases when traversing from maximum to minimum of the spectral weight. Inhomogeneity of spectral weight distribution can be characterized by ratio \\$\Delta A = {{A_{\min }^ - } \mathord{\left/
 {\vphantom {{A_{\min }^ - } {A_{\max }^ + }}} \right.
 \kern-\nulldelimiterspace} {A_{\max }^ + }}$. Parameter of inhomogeneity for the hole concentration $x = 0.03$ is equal to $\Delta A\left( {x = 0.03} \right) = 0.381$ (Fig.~\ref{fig:Inhom}c).

Increasing hole concentration shifts chemical potential deeper to VB, reconstructs band structure and redistributes spectral weight. VB is modified stronger than CB. In the CB doping results in the flat band formation at point ${\bf{k}} = \left( {\pi ,0} \right)$. VB transformation is associated with increase of quasiparticle energy at ${\bf{k}} = \left( {\pi ,\pi } \right)$ (Fig.~\ref{fig:BandStr}c) which is consequence of spin-spin correlation function damping with doping.  Formation of local energy maximum in the direction ${\bf{k}} = \left( {\pi ,0} \right)$ - ${\bf{k}} = \left( {\pi ,\pi } \right)$ accompanies the VB reconstruction. Hole pocket becomes larger with doping (Fig.~\ref{fig:FS}b), spectral weight inhomogeneity increases (Fig.~\ref{fig:Inhom}b, red line), parameter $\Delta A\left( {x = 0.1} \right) = 0.215$. Reduced intensity of quasiparticle excitations at low-intensity FC may be a reason of its absence in the ARPES measured Fermi surfaces. In fact arc can be high-intensity part of hole pocket. Wherein with doping absolute value of maximal spectral weight in the high-intensity FC becomes larger and absolute value of minimal spectral weight in the low-intensity FC becomes smaller (Fig.~\ref{fig:Inhom}a,b).

 When chemical potential at ${x_{c1}} = 0.1601$ touches local maximum of dispersion surface at k-point in the directions $\left( {\pi ,0} \right) - \left( {\pi ,\pi } \right)$ and $\left( {0,\pi } \right) - \left( {\pi ,\pi } \right)$ first quantum phase transition (QPT) of the Lifshits type occurs. This QPT is accompanied by closing of hole pockets (Fig.~\ref{fig:FS}c) with their transformation into two large contours, bigger one is hole contour, smaller one has electron type (Fig.~\ref{fig:FS}d). Spectral weight is very different in the hole and electron contours, parameter of inhomogeneity $\Delta A\left( {x = 0.2} \right) = 0.177$. However intensity gradient along each contour individually begins to decrease with doping after first QPT (Fig.~\ref{fig:Inhom}a,b, green and purple lines). It is seen from Fig.~\ref{fig:Inhom}b that spectral weight of the low-intensity FC is almost angle and momentum independent at $x = 0.2$ and $x = 0.25$. Small ratio $\Delta A$ may be the reason why small sectors of the FC cannot be found in the ARPES~\cite{Norman1998,ShenKM2005}, while quantum oscillation experiments can determine the area of a closed contour~\cite{Doiron2007,Yelland2008}. Further doping leads to expansion of the hole contour and the reduction of electron contour. Second QPT occurs when chemical potential crosses local minimum of VB at point ${\bf{k}} = \left( {\pi ,\pi } \right)$ at ${x_{c2}} \approx 0.283$ (Fig.~\ref{fig:FS}e): electron contour disappears and only hole contour remains (Fig.~\ref{fig:FS}f). Thus Fermi surface of the overdoped system ($x = 0.3$) is slightly inhomogeneous hole contour around ${\bf{k}} = \left( {\pi ,\pi } \right)$ (Fig.~\ref{fig:FS}f). Similar transformations of the electronic structure was obtained within t-J-model in the work~\cite{Barabanov2001} and the Hubbard model~\cite{Plakida2007,Avella2014}. The critical values for the Lifshitz transition obtained in~\cite{KorOvch2007} for t-J model are ${x_{c1}} = 0.15$ and ${x_{c2}} = 0.24$. The inhomogeneous distribution of the spectral weight over dispersion surface has been obtained also in the Hubbard model~\cite{Plakida2007,Avella2014}.

\section{Strain dependence of the electronic structure \label{Strain dependence}}

\begin{figure}
\center
\includegraphics[width=0.98\linewidth]{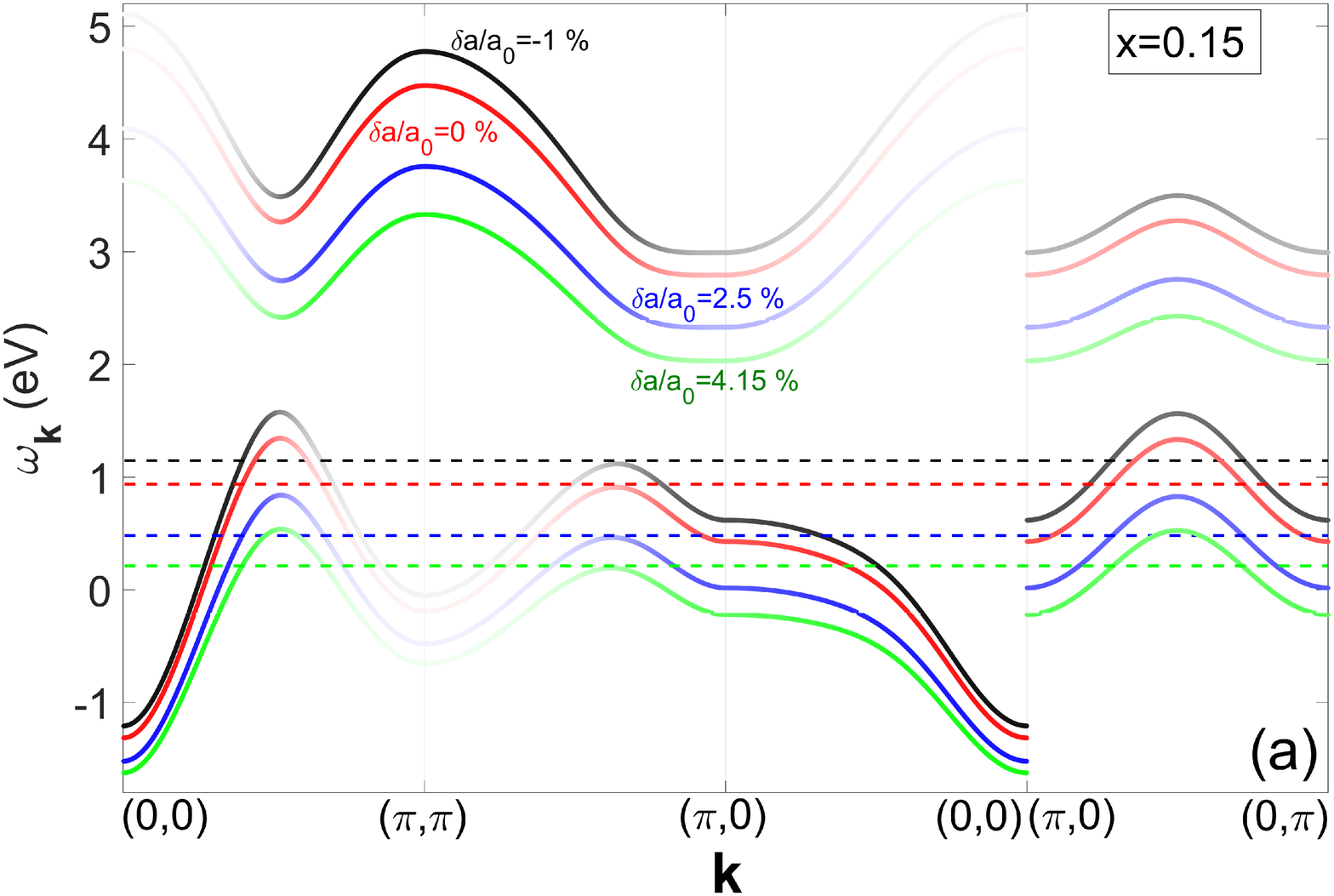}
\includegraphics[width=0.8\linewidth]{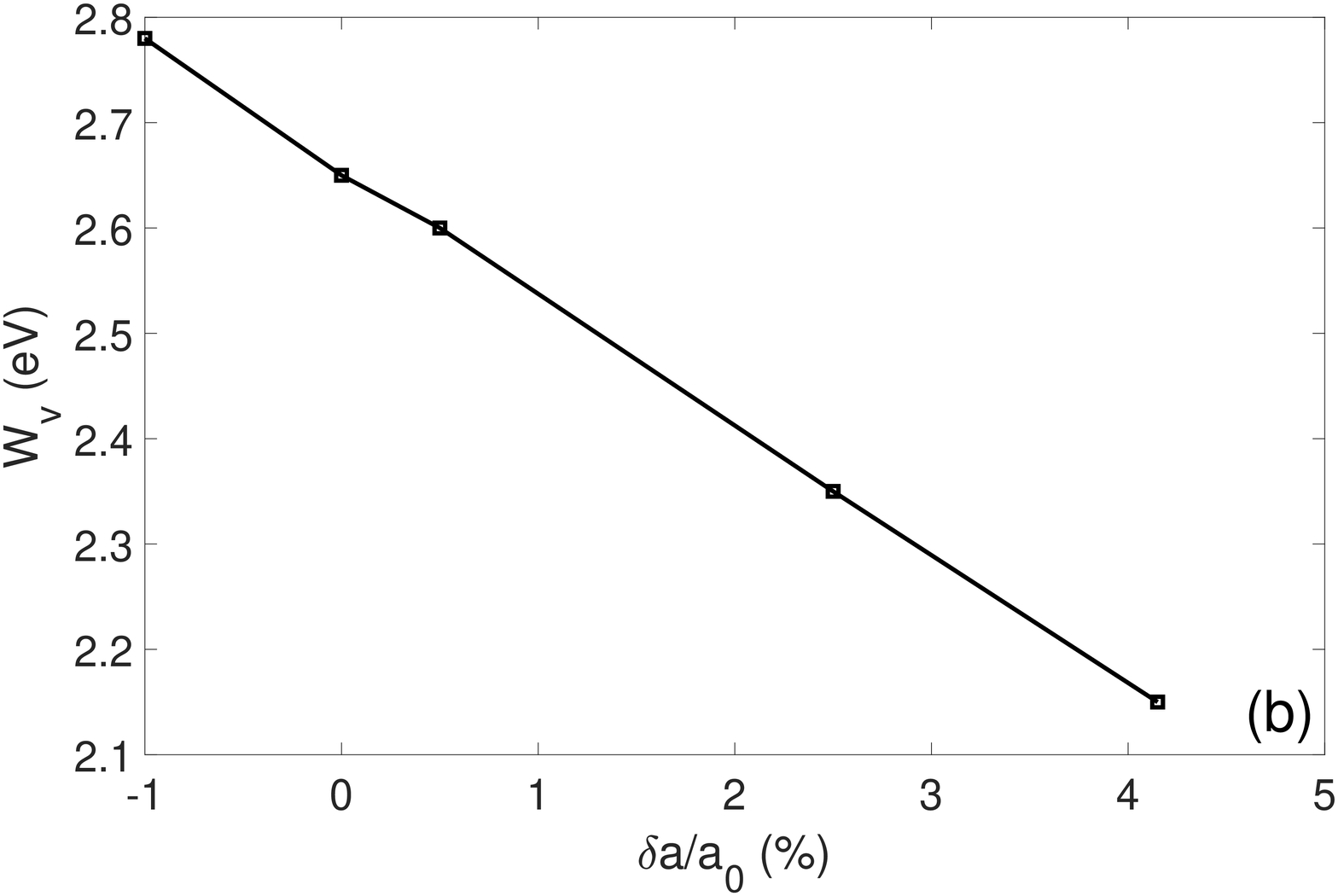}
\caption{\label{fig:Strain} (Color online) Strain dependence of the (a) band structure of quasiparticle excitations and (b) the valence band width ${W_v}$ at doping $x = 0.15$. Dashed lines indicate the Fermi levels}
\end{figure}

\begin{figure}
\center
\includegraphics[width=0.98\linewidth]{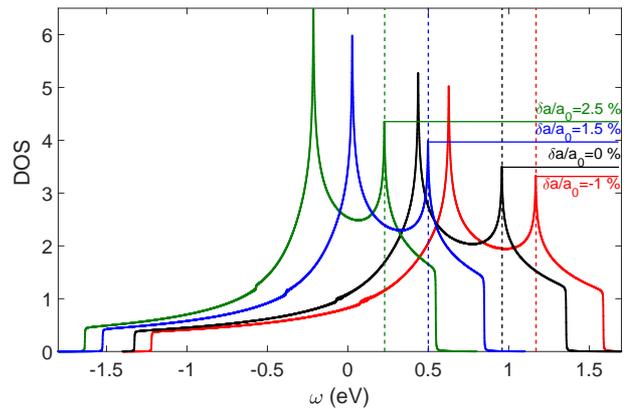}
\caption{\label{fig:DOS} (Color online) Density of states at concentrations of doped holes ${x_{c1}}$ corresponding to first quantum phase transition for given variation of lattice parameter $\delta a/a_0$. Dashed lines indicate the Fermi levels}
\end{figure}

\begin{table*} [htb]
\center
\begin{tabular}{c@{\hspace{3mm}}c@{\hspace{10mm}}*{12}{c@{\hspace{6mm}}}}
Compound & ${T_{c\max }},\,K$ & $\delta a/a_0, \%$ & ${x_{c1}}$ & ${x_{c2}}$ & $J\,\left( {eV} \right)$ & ${N_{\max }}\left( {{\mu _{c1}}} \right)$ & $J{N_{\max }}\left( {{\mu _{c1}}} \right)$ \\
\hline
  -- & -- &	$-1$ &	0.1604	& 0.285	& 0.168	& 3.32956	& 0.5594 \\
  La$_{2+x}$Sr$_x$CuO$_4$ & 40	& $0$	& 0.1601	& 0.2828	& 0.156	& 3.499	& 0.5458 \\
  -- & --	&$0.5$	& 0.16	& 0.2815	& 0.15	& 3.5862	& 0.5379 \\
  Bi$_2$Sr$_2$CaCu$_2$O$_{8+x}$ & 82	& $1.5$	& 0.1599	& 0.28	& 0.14	& 3.7355	& 0.5230 \\
  HgBa$_2$Ca$_2$Cu$_3$O$_{8+\delta}$ & 135	& $2.5$	& 0.1594	& 0.277	& 0.127	& 3.9886	& 0.5066 \\
  HgBa$_2$CaCu$_2$O$_{6+\delta}$ & 127	& $2.7$	& --	& --	& --	& --	& -- \\
  HgBa$_2$CuO$_{4+\delta}$ & 97	& $3.1$	& --	& --	& --	& --	& -- \\
  -- & --& $4.15$	& 0.1589	&	0.2735	&	0.112	&	4.3413	&	0.4862 \\
\end{tabular}
\caption{Concentrations of the first and second quantum phase transitions (${x_{c1}}$ and ${x_{c2}}$), superexchange parameter ($J$), DOS maximum at chemical potential for concentration ${x_{c1}}$ (${N_{\max }}\left( {{\mu _{c1}}} \right)$) and their product ($J{N_{\max }}\left( {{\mu _{c1}}} \right)$) at different values of the strain $\delta a/a_0$. First row shows HTSC compound with lattice parameter $a\left( b \right)$ corresponding to a given strain, experimental ${T_{c\max }}$ of the related compound is in second row. }
\label{QPTTc}
\end{table*}

There are two main effects of lattice parameter increasing on band structure: (I) energy shift of the VB and CB; (II) their bandwidths decrease. Bands are shifted to lower energies (Fig.~\ref{fig:Strain}a). This shift is not rigid, its value depends on wave vector ${\bf{k}}$. Transformation of the band structure with lattice parameter changing has the same character for entire range of hole concentrations   under study (from 0.03 to 0.3).

Evolution of the Fermi contour with doping at different lattice parameters is qualitatively the same as at ${{\delta a} \mathord{\left/
 {\vphantom {{\delta a} {{a_0}}}} \right.
 \kern-\nulldelimiterspace} {{a_0}}} = 0 \%$. But concentrations of doped holes at which two QPTs occur are slightly shifted to smaller values with lattice parameter increasing (Table~\ref{QPTTc}). Chemical potential is in the maximum of density of states (DOS) at hole concentration ${x_{c1}}$ of the first QPT. Therefore it is believed that ${T_c}$ maximum is determined by the critical concentration ${x_{c1}}$. The magnitude of DOS maximum ${N_{\max }}\left( {{\mu _{c1}}} \right)$ increases with lattice parameter increasing (Fig.~\ref{fig:DOS}, Table~\ref{QPTTc}). Moreover the larger lattice parameter the higher DOS at all energy regions. This fact results from that fixed number of states (${n_h} = 1 + x$) is distributed over smaller energy interval for the larger lattice parameter since bandwidth decreases with strain increasing (Fig.~\ref{fig:Strain}b). Strain dependence of superexchange interaction $J$ demonstrates behavior inversed to ${N_{\max }}\left( {{\mu _{c1}}} \right)$: $J$ noticeably increases with a decreasing ${{\delta a} \mathord{\left/
 {\vphantom {{\delta a} {{a_0}}}} \right.
 \kern-\nulldelimiterspace} {{a_0}}}$ (Table~\ref{QPTTc}) since this type of strain is the most effective way to enhance superexchange in LCO~\cite{GavPchNekOvch2016}. This fact is easy to understand because $J$ is proportional to hopping integrals ${t_{pd}}$ and ${t_{pp}}$  which decrease with in-plane lattice parameter growing. Superexchange constant was calculated in the framework of LDA+GTB approach, where superexchange interaction derived in analytical form~\cite{GavPolOvch2017} due to the many-electron approach based on the X-operator representation~\cite{Hubbard1965} and technique of projection operators~\cite{Chao1977} generalized for arbitrary quasiparticle energy spectra of Mott-Hubbard insulator. Effective pairing constant of exchange interaction $J{N_{\max }}\left( {{\mu _{c1}}} \right)$ entering into ${T_c}$ equation also decreases with in-plane lattice parameter increasing. Therefore it should be expected that calculated ${T_c}$ will grow with lattice parameter $a\left( b \right)$ reducing in agreement to experimental data on uniaxial pressure in La$_{2-x}$Sr$_x$CuO$_4$~\cite{Gugenberger1994} and Bi$_2$Sr$_2$CaCu$_2$O$_{8+x}$~\cite{Meingast1996}. Thus it is seen that strain dependence of ${T_c}$ is dictated by behavior of $J$.

\section{Conclusion \label{Summary}}

Evolution of electronic structure of CuO$_6$-octahedra layer with doping is obtained for different Cu-O distances $a\left( b \right)$. With hole doping Fermi contour is transformed from four small hole pockets with inhomogeneous spectral weight to large hole and electron contours and then only large hole contour with almost homogeneous spectral weight distribution remains. Inhomogeneity of the spectral weight on the opposite sides of the hole pockets relative to the boundary of the antiferromagnetic Brillouin zone increases with hole doping. Low spectral weight on the one side potentially can be reason of its absence in the ARPES. Changes of the electronic structure with lattice parameter $a\left( b \right)$ varying are quantitative but not qualitative. Band structure is shifted to smaller energies, width of the valence and conductivity bands shrinks, concentrations of the first and second QPT are shifted with strain increasing. We obtain strain dependence of the characteristics determining superconducting temperature ${T_c}$ within the mean-field theory with magnetic mechanism of pairing, DOS and antiferromagnetic exchange constant $J$. The singularity of the DOS at the chemical potential energy at concentration of doped holes corresponding to first QPT ${N_{\max }}\left( {{\mu _{c1}}} \right)$ demonstrates monotonic growth with lattice parameter $a\left( b \right)$ increasing. Exchange constant $J$ monotonically decreases with strain increasing as well as resulting effective exchange constant $J{N_{\max }}\left( {{\mu _{c1}}} \right)$  in the ${T_c}$ equation. Strain dependences of both characteristics $J$ and ${N_{\max }}\left( {{\mu _{c1}}} \right)$ are managed by one type of parameters, hopping integrals. Superexchange interaction is directly determined by hopping integral and DOS depends on bandwidth which is also defined by hopping integrals. Growth of DOS and damping of $J$ are qualitative effects of the lattice parameter $a\left( b \right)$ increasing, in the general case behavior of the product $J{N_{\max }}\left( {{\mu _{c1}}} \right)$ depends on rate of change of each of the characteristics $J$ and ${N_{\max }}\left( {{\mu _{c1}}} \right)$ with lattice parameter $a\left( b \right)$ varying. For example slight increase in the growth rate of the function ${N_{\max }}\left( {{\mu _{c1}}} \right)$ of the lattice parameter $a\left( b \right)$ in comparison with that shown in the Table~\ref{QPTTc} results in the nonmonotonic dependence of $J{N_{\max }}\left( {{\mu _{c1}}} \right)$ on lattice parameter. There is nonmonotonic dependence of experimental ${T_{c\max }}$ on the lattice parameter $a\left( b \right)$ among all compounds of different cuprate families (Table~\ref{QPTTc}, second row). Therefore one can expect that different material dependent parameters such as bilayer splitting, chemical and electronic inhomogeneity and others play important role in the formation of the electronic structure and DOS.

Finally we have discussed the effect of a variable lattice strain on the unconventional "Lifshitz transitions" appearing in the strongly correlated electronic structure of the CuO$_2$  as a function of doping. "Lifshitz transitions" are recently becoming a hot topic in the models of high-temperature superconductivity since there is growing evidence for the emergence of the "superconducting domes" in different high-temperature superconductors tuning the chemical potential near Lifshitz transitions~\cite{Innocenti2010,Innocenti2011,Perali2012,Bianconi2016,Jarlborg2016,Mazziotti2017}.

\begin{acknowledgements}
We thank the Presidium RAS program No.12 "Fundamental problems of high-temperature superconductivity" for the financial support under the project 0356-2018-0063. The reported study was funded by the Russian Foundation for Basic Research, Government of Krasnoyarsk Territory and Krasnoyarsk Regional Fund of Science according to the research project:"Features of electron-phonon coupling in high-temperature superconductors with strong electronic correlations" No. 18-42-240017. This work was done under the State contract (FASO) No. 0389-2014-0001 and supported in part by RFBR grants No. 17-02-00015 and 18-02-00281.
\end{acknowledgements}


\begin{thebibliography}{}

\bibitem{Arrouy1996} Arrouy,~F., Locquet,~J.-P., Williams,~E.J.,  M\"{a}chler,~E., Berger,~R., Gerber,~C.,  Monroux,~C., Grenier,~J.-C., Wattiaux,~A.: Growth, microstructure, and electrochemical oxidation of MBE-grown c-axis La$_2$CuO$_4$ thin films. Phys. Rev. B {\bf 54}, 7512-7520 (1996)

\bibitem{Sato1997} Sato,~H., Naito,~M.: Increase in the superconducting transition temperature by anisotropic strain effect in (001) La$_{1.85}$Sr$_{0.15}$CuO$_4$ thin films on LaSrAlO$_4$ substrates. Physica C {\bf 274}, 221-226 (1997)

\bibitem{Lockuet1998} Locquet,~J.-P., Perret,~J., Fompeyrine,~J., M\"{a}chler,~E., Seo,~J.W., Van Tendeloo,~G.: Doubling the critical temperature of La$_{1.9}$Sr$_{0.1}$CuO$_4$ using epitaxial strain. Nature {\bf 394}, 453-456 (1998)

\bibitem{Sato2000} Sato,~H., Tsukada,~A., Naito,~M., Matsuda,~A.: La$_{2-x}$Sr$_x$CuO$_y$ epitaxial thin films (x=0 to 2): Structure, strain and superconductivity. Phys. Rev. B {\bf 61}, 12447-12456 (2000)

\bibitem{Attfield1998} Attfield,~J.P., Kharlanov,~A.L., McAllister,~J.A.: Cation effects in doped La$_2$CuO$_4$ superconductors. Nature {\bf 394}, 157-159 (1998)

\bibitem{AgrBB2003} Agrestini,~S., Saini,~N.L., Bianconi,~G., Bianconi,~A.: The strain of CuO$_2$ lattice: the second variable for the phase diagram of cuprate perovskites. J. Phys. A: Math. Gen. {\bf 36}, 9133-9142 (2003)

\bibitem{BiancSaiAgr2000} Bianconi,~A., Saini,~N.L., Agrestini,~S., Di Castro,~D., Bianconi,~G.: The strain quantum critical point for superstripes in the phase diagram of all cuprate perovskites. Int. J. Mod. Phys. B {\bf 14}, 3342-3355 (2000)

\bibitem{Kusmartsev2000} Kusmartsev,~F.V., Di Castro,~D., Bianconi,~G., Bianconi,~A.: Transformation of strings into an inhomogeneous phase of stripes and itinerant carriers. Physics Letters A {\bf 275}, 118-123 (2000)

\bibitem{BiancCastBianc2000} Bianconi,~A., Di Castro,~D., Bianconi,~G., Pifferi,~A., Saini,~N.L., Chou,~F.C., Johnston,~D.C., Colapietro, M.: Coexistence of stripes and superconductivity: $T_c$ amplification in a superlattice of superconducting stripes. Physica C {\bf 341-348}, 1719-1722 (2000)

\bibitem{Buchner1994} B\"{u}chner,~B., Breuer,~M., Freimuth,~A., Kampf,~A.P.: Critical buckling for the disappearance of superconductivity in rare-earth-doped La$_{2-x}$Sr$_x$CuO$_4$. Phys. Rev. Lett. {\bf 73}, 1841-1844 (1994)

\bibitem{Anderson1987} Anderson,~P.W.: The resonating valence bond state in La$_2$CuO$_4$
and superconductivity. Science {\bf 235}, 1196-1198 (1987)

\bibitem{Bianconi1987} Bianconi,~A., Congiu Castellano,~A., De Santis,~M., Rudolf,~P.,  Lagarde,~P., Flank,~A.M., Marcelli,~A.: L$_{2,3}$ xanes of the high $T_c$ superconductor YBa$_2$Cu$_3$O$_{\approx7}$ with variable oxygen content. Solid State Commun. {\bf 63}, 1009-1013 (1987).

\bibitem{Emery1987} Emery,~V.J.: Theory of High-$T_c$ Superconductivity in Oxides. Phys. Rev. Lett. {\bf 58}, 2794-2797 (1987); Varma,~C.M., Schmitt-Rink,~S., Abrahams,~E.: Charge transfer excitations and superconductivity in "ionic" metals. Solid State Commun. {\bf 62}, 681-685 (1987)

\bibitem{KorGavOvch2004} Korshunov,~M.M., Gavrichkov,~V.A., Ovchinnikov,~S.G., Pchelkina,~Z.V., Nekrasov,~I.A., Korotin,~M.A., Anisimov,~V.I.: Parameters of the effective singlet-triplet model for band structure of high-$T_c$ cuprates by alternative approaches. JETP {\bf 99}, 559-565 (2004)

\bibitem{KorGavOvch2005} Korshunov,~M.M., Gavrichkov,~V.A., Ovchinnikov,~S.G., Nekrasov,~I.A., Pchelkina,~Z.V., Anisimov,~V.I.: Hybrid LDA and generalized tight-binding method for electronic structure calculations of strongly correlated electron systems. Phys. Rev. B. {\bf 72}, 165104 (2005)

\bibitem{Bianconi1996} Bianconi,~A., Saini,~N.L., Lanzara,~A., Missori,~M., Rossetti,~T., Oyanagi,~H., Yamaguchi,~H., Oka,~K., Ito,~T.: Determination of the local lattice distortions in the CuO$_2$ plane of La$_{1.85}$Sr$_{0.15}$CuO$_4$. Phys. Rev. Lett. {\bf 76}, 3412-3415 (1996)

\bibitem{Andersen1975} Andersen,~O.K.: Linear methods in band theory. Phys. Rev. B {\bf 12}, 3060-3083 (1975)

\bibitem{GunnJepAnd1983} Gunnarsson,~O., Jepsen,~O., Andersen,~O.K.: Self-consistent impurity calculations in the atomic-spheres approximation. Phys. Rev. B {\bf 27}, 7144-7168 (1983)

\bibitem{AndJep1984} Andersen,~O.K., Jepsen,~O.: Explicit, first-principles tight-binding theory. Phys. Rev. Lett. {\bf 53}, 2571-2574 (1984)

\bibitem{Amulet} http://amulet-code.org/tb-lmto-2

\bibitem{GunnAndJepZaa1989} Gunnarsson,~O., Andersen,~O.K., Jepsen,~O., Zaanen,~J.: Density-functional calculation of the parameters in the Anderson model: Application to Mn in CdTe. Phys. Rev. B {\bf 39}, 1708-1722 (1989)

\bibitem{AnGunn1991} Anisimov,~V.I., Gunnarsson,~O.: Density-functional calculation of effective Coulomb interactions in metals. Phys. Rev. B {\bf 43}, 7570-7574 (1991)

\bibitem{OvchSan1989} Ovchinnikov,~S.G., Sandalov,~I.S.: The band structure of strong-correlated electrons in La$_{2-x}$Sr$_x$CuO$_4$ and YBa$_2$Cu$_3$O$_{7-y}$. Physica C {\bf 161}, 607-617 (1989)

\bibitem{GavOvchBorGor2000} Gavrichkov,~V.A., Ovchinnikov,~S.G., Borisov,~A.A., Goryachev,~E.G.: Evolution of the band structure of quasiparticles with doping in copper oxides on the basis of a generalized tight-binding method. JETP {\bf 91}, 369-383 (2000)

\bibitem{KuzOvch1977} Kuz'min,~E.V., Ovchinnikov,~S.G.: Electron correlations in a Hubbard antiferromagnetic semiconductor. Weak coupling. Teor. Mat. Fiz. {\bf 31}, 523-531 (1977)

\bibitem{OvchVal2004} Ovchinnikov,~S.G., Val'kov,~V.V.: Hubbard Operators in the Theory of Strongly Correlated Electrons. Imperial College Press, London-Singapure (2004)

\bibitem{PlAnAdAd2003} Plakida,~N.M., Anton,~L., Adam,~S., Adam,~G.: Exchange and spin-fluctuation mechanisms of superconductivity in cuprates. JETP  {\bf 97}, 331-342 (2003)

\bibitem{KorOvch2007} Korshunov,~M., Ovchinnikov,~S.: Doping-dependent evolution of low-energy excitations and quantum phase transitions within an effective model for high-$T_c$ copper oxides. Eur. Phys. J. B. {\bf 57}, 271-278 (2007)

\bibitem{MakOvch2015} Makarov,~I.A., Ovchinnikov,~S.G.: Temperature dependence of the electronic structure of La$_2$CuO$_4$ in the multielectron LDA+GTB approach. JETP {\bf 148}, 526-534 (2015)

\bibitem{NekKuchPchSad2007}  Nekrasov,~I.A., Kuchinskii,~E.Z., Pchelkina,~Z.V., Sadovskii,~M.V.: Pseudogap in normal underdoped phase of Bi2212: LDA+DMFT+$\Sigma_k$. Physica C {\bf 460-462}, 997-999 (2007)

\bibitem{Nekrasov2009} Nekrasov,~I.A., Pavlov,~N.S., Kuchinskii,~E.Z., Sadovskii,~M.V., Pchelkina,~Z.V., Zabolotnyy,~V.B. Geck,~J., B\"{u}chner,~B., Borisenko,~S.V., Inosov,~D.S., Kordyuk,~A.A., Lambacher,~M., Erb,~A.: Electronic structure of Pr$_{2-x}$Ce$_x$CuO$_4$ studied via ARPES and LDA+DMFT+$\Sigma_k$. Phys. Rev. B {\bf 80}, 140510R (2009)

\bibitem{Norman1998} Norman,~M., Ding,~H., Randeria,~M., Campuzano,~J.C., Yokoya,~T., Takeuchik,~T., Takahashi,~T., Mochiku,~T., Kadowaki,~K., Guptasarma,~P., Hinks,~D.G.: Destruction of the Fermi surface in underdoped high-$T_c$ superconductors. Nature {\bf 392}, 157-160 (1998)

\bibitem{ShenKM2005} Shen,~K.M., Ronning,~F., Lu,~D.H., Baumberger,~F., Ingle,~N.J.C., Lee,~W.S.,  Meevasana,~W., Kohsaka,~Y., Azuma,~M., Takano,~M., Takagi,~H., Shen,~Z.-X.: Nodal quasiparticles and antinodal charge ordering in Ca$_{2-x}$Na$_x$CuO$_2$Cl$_2$. Science {\bf 307}, 901-904 (2005)

\bibitem{Doiron2007} Doiron-Leyraud,~N., Proust,~C., LeBoeuf,~D., Levallois,~J., Bonnemaison,~J.-B., Liang,~R., Bonn,~D.A., Hardy,~W.N., Taillefer,~L.: Quantum oscillations and the Fermi surface in an underdoped high-$T_c$ superconductor. Nature {\bf 447} 565-568 (2007)

\bibitem{Yelland2008} Yelland,~E.A., Singleton,~J., Mielke,~C.H., Harrison,~N., Balakirev,~F.F., Dabrowski,~B., Cooper,~J.R.: Quantum oscillations in the underdoped cuprate YBa$_2$Cu$_4$O$_8$. Phys. Rev. Lett. {\bf 100}, 047003 (2008)

\bibitem{Barabanov2001} Barabanov,~A.F., Kovalev,~A.A., Urazaev,~O.V., Belemuk,~A.M., Hayn,~R.: Evolution of the Fermi surface of cuprates on the basis of the spin-polaron approach. JETP {\bf 92}, 677-695 (2001)

\bibitem{Plakida2007} Plakida,~N.M., Oudovenko,~V.S.: Electron spectrum in high-temperature cuprate superconductors. JETP {\bf 104}, 230-244 (2007)

\bibitem{Avella2014} Avella,~A.: Composite operator method analysis of the underdoped cuprates puzzle. Advances in Condensed Matter Physics {\bf 2014}, 515698 (2014)

\bibitem{GavPchNekOvch2016} Gavrichkov,~V.A., Pchelkina,~Z.V., Nekrasov,~I.A., Ovchinnikov,~S.G.: Pressure effect on the energy structure and superexchange interaction in undoped orthorhombic La$_2$CuO$_4$. Int. J. Mod. Phys. B {\bf 30}, 1650180 (2016)

\bibitem{GavPolOvch2017} Gavrichkov,~V.A., Polukeev,~S.I., Ovchinnikov,~S.G.: Contribution from optically excited many-electron states to the superexchange interaction in Mott-Hubbard insulators. Phys Rev B {\bf 95}, 144424 (2017)

\bibitem{Hubbard1965} Hubbard,~J.: Electron correlations in narrow energy bands. IV. The atomic representation. Proc. R. Soc. A {\bf 285}, 542-560 (1965)

\bibitem{Chao1977} Chao,~K.A., Spalek,~J., Ole\'{s},~A.M.: J. Phys. C: Kinetic exchange interaction in a narrow S-band. Solid State Phys. {\bf 10}, L271-L276 (1977)

\bibitem{Gugenberger1994} Gugenberger,~F., Meingast,~C., Roth,~G., Grube,~K., Breit,~V., Weber,~T., Wuhl,~H., Uchida,~S., Nakamura,~Y.: Uniaxial pressure dependence of $T_c$ from high-resolution dilatometry of untwinned La$_{2-x}$Sr$_x$CuO$_4$ single crystals. Phys. Rev. B {\bf 49}, 13137-13142 (1994)

\bibitem{Meingast1996} Meingast,~C., Junod,~A., Walker,~E.: Superconducting fluctuations and uniaxial-pressure dependence of $T_c$ of a Bi$_2$Sr$_2$CaCu$_2$O$_{8+x}$ single crystal
from high-resolution thermal expansion. Physica (Amsterdam) {\bf 272C}, 106-114 (1996)

\bibitem{Innocenti2010} Innocenti,~D., Poccia,~N., Ricci,~A., Valletta,~A., Caprara,~S., Perali,~A., Bianconi,~A.: Resonant and crossover phenomena in a multiband superconductor: tuning the chemical potential near a band edge. Phys. Rev. B {\bf 82}, 184528 (2010)

\bibitem{Innocenti2011} Innocenti,~D., Valletta,~A., Bianconi,~A.: Shape resonance at a Lifshitz transition for high temperature superconductivity in multiband superconductors. J. Supercond. Nov. Magn. {\bf 24}, 1137-1143 (2011)

\bibitem{Perali2012} Perali,~A., Innocenti,~D., Valletta,~A., Bianconi,~A.: Anomalous isotope effect near a 2.5 Lifshitz transition in a multi-band multi-condensate superconductor made of a superlattice of stripes. Supercond. Sci. Technol. {\bf 25}, 124002 (2012)

\bibitem{Bianconi2016} Bianconi,~A., Jarlborg,~T.: Superconductivity above the lowest Earth temperature in pressurized sulfur hydride. EPL {\bf 112}, 37001 (2015)

\bibitem{Jarlborg2016} Jarlborg,~T., Bianconi,~A.: Breakdown of the Migdal approximation at Lifshitz transitions with giant zero-point motion in the H$_3$S superconductor. Scientific Reports {\bf 6}, 24816 (2016)

\bibitem{Mazziotti2017} Mazziotti,~M.V., Valletta,~A., Campi,~G., Innocenti,~D., Perali,~A., Bianconi,~A.: Possible Fano resonance for high-$T_c$ multi-gap superconductivity in p-Terphenyl doped by K at the Lifshitz transition. EPL {\bf 118}, 37003 (2017)


\end{thebibliography}


\end{document}